\documentclass[showpacs,preprintnumbers]{revtex4}
\usepackage{amssymb}
\usepackage{amsmath}
\usepackage{graphicx}
\usepackage{dcolumn}
\usepackage{bm}

\begin{document}

\title{Self-organization effects and light amplification of collective
atomic recoil motion in a harmonic trap}

\author{Lin Zhang}
\email{zhanglin.cn@mail.edu.cn} \affiliation{Nonlinear Quantum
Physics Research Institute, Baoji university of arts and science,
Baoji 721007, P. R. China}

\author{G.J.Yang}
\author{L.X.Xia}
\affiliation{Physics Department, Beijing Normal University, Beijing 100875, P. R. China}
\date{\today}

\begin{abstract}
Self-organization effects related to light amplification in the
collective atomic recoil laser system with the driven atoms
confined in a harmonic trap are investigated further. In the
dispersive parametric region, our study reveals that the
spontaneously formed structures in the phase space contributes an
important role to the light amplification of the probe field under
the atomic motion being modified by the trap.
\end{abstract}

\maketitle


\section{Introduction and simulation results}

Optical amplification in a strongly driven sodium atoms system
without population inversions has been observed in the experiment
\cite{Mollow} where the atomic recoil effect and Doppler shift
were purposely avoided by a specific systematic configuration.
When the pumping field is off-detuned from the atomic transition
frequency $\omega_0$, the experimental pump-probe gain spectrum
exhibits a normal absorption-gain profile similar to that of
Fig.1(a) which is drawn from a direct simulation and has been
obtained by Bonifacio \emph{et al} in their original work
\cite{Boni}.
\begin{figure}[htp]%
\includegraphics[scale=0.42]{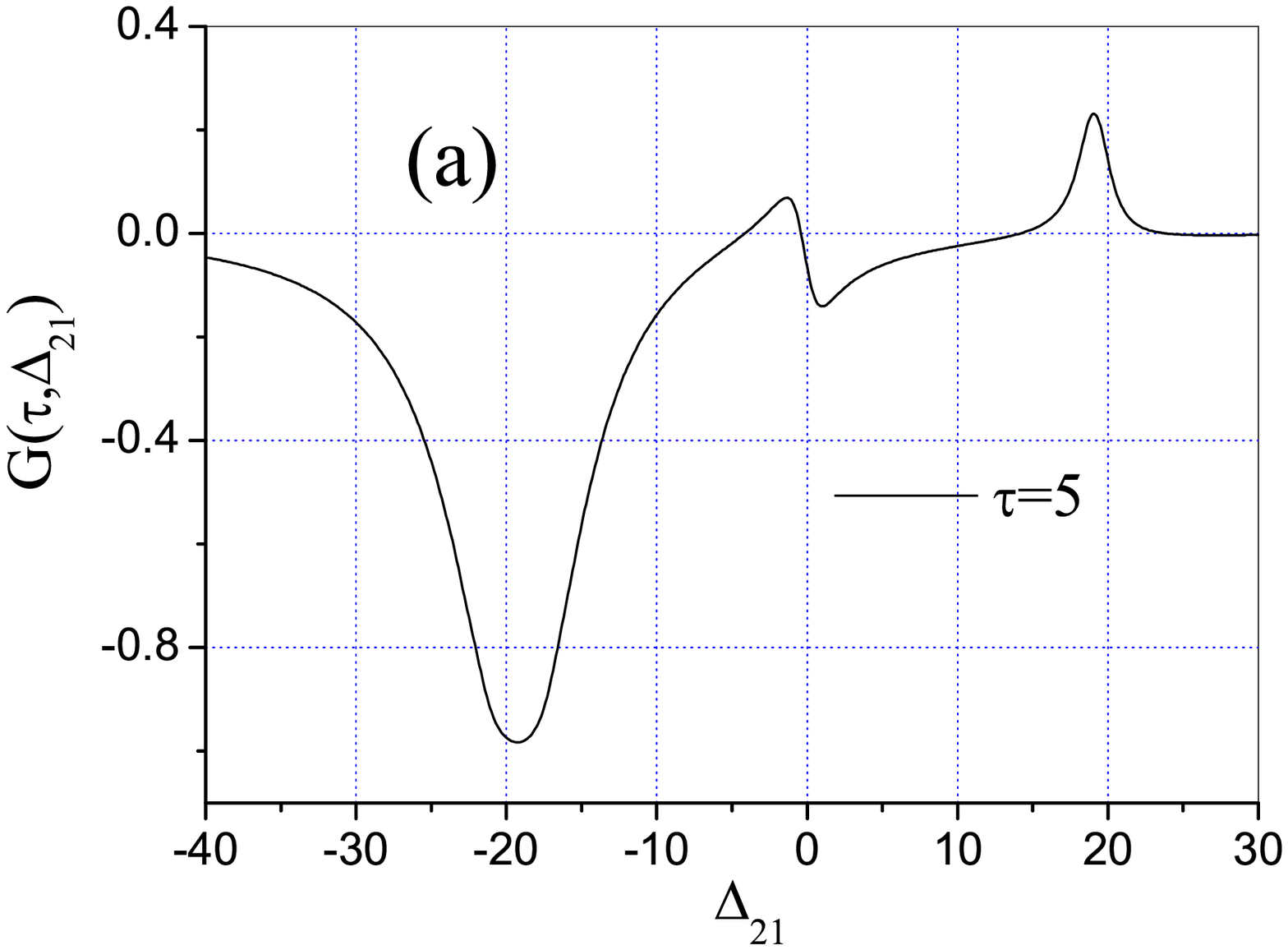} %
\includegraphics[scale=0.42]{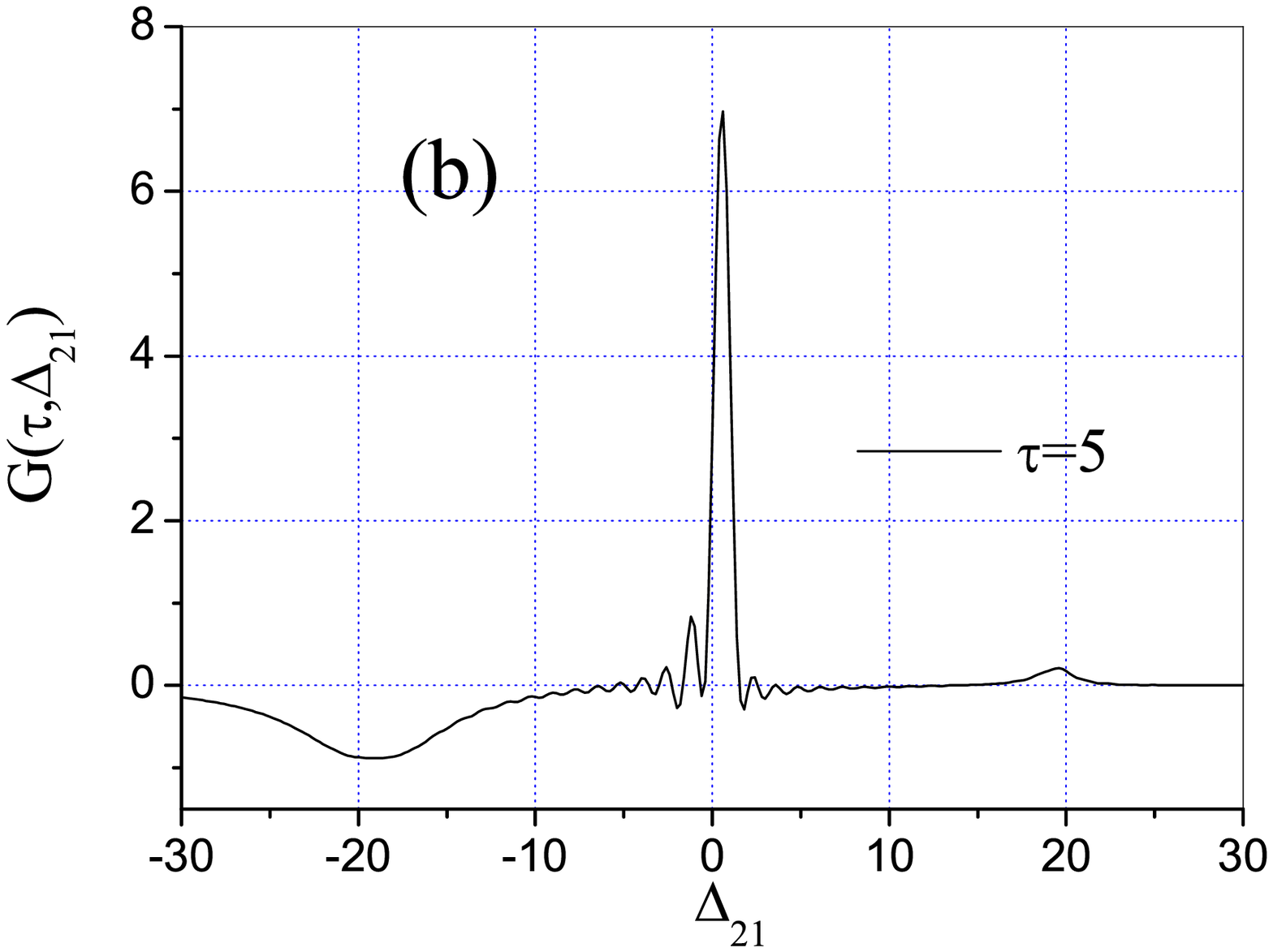}
\caption{Gain profile $G(\protect\tau ,\Delta _{21})$ of the probe
field versus pump-probe detuning $\Delta _{21}$ with parameters
$\protect\nu =0$, $\Gamma =1, \protect\kappa=0.01$, $\protect\rho
=3$, $A_2=2$, $\Delta _{20}=-15$ and $\protect\tau =5$. (a) the
atomic motion is totally removed from the model; (b) the motion
effect is included (traditional CARL).}
\end{figure}
The vertical ordinate of Fig.1 stands for a relative gain of the
probe field and the horizontal ordinate $\Delta_{21}$ denotes the
detuning between the pump and the probe field. However, this
anti-symmetry gain profile around $\Delta_{21}=0$ can be
drastically changed (Fig.1(b)) by the atomic recoil motion which
plays an important role in the amplification mechanism of the
collective atomic recoil laser (CARL) \cite{Boni}. As shown in
Fig.1(b), the gain peak within the traditional Madey region
\cite{Madey} is greatly enhanced, and which has been a signature
of light amplification induced by cooperative atomic recoil
effect. Being modified by the recoil motion and exchanging
energies with one common pump field, the atoms will exhibit some
kinds of self-organization behavior \cite{Domokos} or dynamic
synchronization \cite{Kuramoto} which induces phase transitions in
the relevant system \cite{Perrin}\cite{Cube}. Indeed, in the
traditional CARL system, the atomic density grating as well as the
population inversion grating on the scale of the electromagnetic
wavelength emerges spontaneously during the light amplification
process \cite{Lippi}. However, some theoretical simplicities of
the traditional CARL model, as have been pointed by some authors
\cite{Perrin}, impede a CW gain output of the system. Therefore,
the steady-state CARLs were proposed and have been experimentally
demonstrated \cite{Lippi}\cite{Kruse}, but, with a relative weak
gain. In this paper, we report a result in order to give a more
detailed theoretical analysis to a modified CARL system
\cite{Yang} which produces a CW laser output with a higher gain.
In our previous scheme, the active cold atoms are trapped in a
harmonic potential, which can be realized by loading the atoms
into an optical lattice or an overall magnetic trap, instead of
introducing a light molasses \cite{Kruse}\cite{Robb} or adding a
thermal collision/momentum damping mechanism to the atomic motion
\cite{Perrin}. We suppose the trap to be strong in radial
direction and loose in axial direction (cigar-shaped trap), which
reducing our system effectively to one-dimension. Our modification
by this configuration produces a very different laser gain profile
illustrated by Fig.2 \cite{Yang}.
\begin{figure}[htp]%
\includegraphics[scale=0.5]{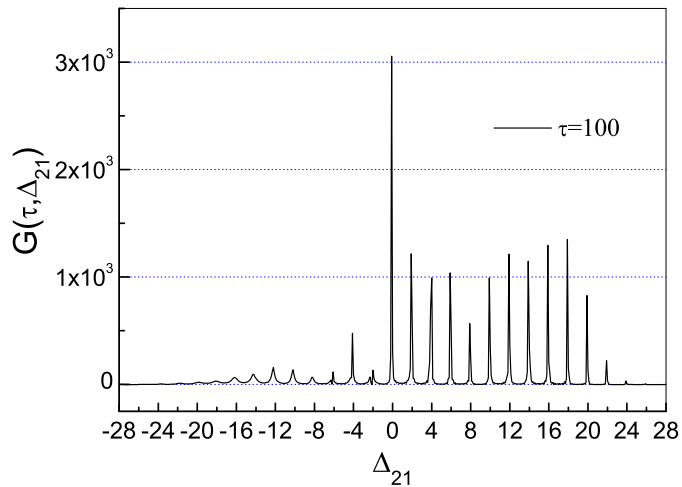}
\caption{Long time gain profile $G(\protect\tau ,\Delta _{21})$ of
the probe field in a trap with frequency of $\protect\nu =2$ and
the time $\protect\tau=100$. The other parameters are the same as
Fig.1 for a convenient comparison.}
\end{figure}
Fig.2 displays a discrete spectral structure which is very similar
to that of a nonlinear Jaynes-Cummings model of one atom in a
high-Q cavity \cite{Vogel}. When the pump-probe detuning satisfies
a Raman condition $ \Delta_{21}=n\nu$, where $n$ is an arbitrary
integer and $\nu$ is the effective axial trapping frequency, a
distinct light amplification process occurs for certain $n$ as
displayed by Fig.2. In this case, a self-organization of all the
atomic oscillators along the trap axis is demonstrated by atomic
distribution pattern and atomic orbit bunching in the phase space
in this paper. The following contains our model with a more
rigorous analysis of all the above interesting results in a
framework of semiclassical method.

\section{Simple model and gain spectrum analysis}

The relevant system is supposed to be enclosed in a ring cavity
\cite{11} and the operating medium is composed of a collection of
$N$ two-level atoms of mass $m$, moving along the axial direction
$z$ of a trap described by
$V_{trap}=\frac{1}{2}m[{\nu_{\perp}}^2(x^2+y^2)+{\nu_z}^2 z^2]$.
The atoms are driven by a strong far off-detuned pump field
propagating along $z$ and probed by a counterpropagating weak
probe field. If we set $\nu_{\perp}\gg \nu_z$, the radial trapping
mode will be hardly excited in the recoil process and a
one-dimensional trap, $V^{\prime}=\frac{1}{2}m {\nu_z}^2 z^2$,
should be a promising approximation. With above simplification,
the explicit form of the Hamiltonian of this system is
\begin{eqnarray}  \label{eq1}
\widehat{H} &=&\sum_{j=1}^N\left( \frac{\widehat{p}_j^2}{2m}+\frac
12m\nu_z^2 \widehat{z}_j^2\right)+\hbar \omega
_0\sum_{j=1}^N\widehat{\sigma }_{zj} +\hbar \omega
_1\widehat{a}_1^{\dagger }\widehat{a}_1+\hbar \omega _2\hat{a}
_2^{\dagger }\hat{a}_2  \notag \\
&&+i\hbar \left( g_1\widehat{a}_1^{\dagger
}\sum_{j=1}^N\widehat{\sigma }
_j^{-}e^{-ik_1\widehat{z}_j}+g_2\widehat{a}_2^{\dagger
}\sum_{j=1}^N\widehat{ \sigma
}_j^{-}e^{+ik_2\widehat{z}_j}-\text{H.c.}\right) ,
\end{eqnarray}
where the operators $\widehat{p}_j$ and $\widehat{z}_j$ are the
canonically conjugate momentum and center-of-mass position of the
$j$th atom, $\widehat{ \sigma }_j^{\mp }$ and $\widehat{\sigma
}_{zj}$ denote the usual spin operators of the atomic internal
degrees of freedom with a transition frequency of $\omega_0$. The
probe field and the pump field with the carrier frequencies
$\omega _1$ and $\omega _2$ are denoted by the photon creation
operators $\widehat{a}_1^{\dagger }$ and $\widehat{a}_2^{\dagger
}$, respectively. The coupling constant between the atom and the
$i$th optical field is given by $g_i$ ($i=1,2$). Following the
conventional commutation relations of all above operators, the
Heisenberg equations for the expectations of the relevant
operators under the semiclassical approximation are
\begin{eqnarray}  \label{eq2}
\frac{d\theta _{j}}{d\tau } &=&p_{j} ,  \notag \\
\frac{dp_{j}}{d\tau } &=&-\nu ^{2}\theta _{j}-A_{1}^{\ast }\sigma
_{j}e^{-i\theta _{j}}-A_{1}\sigma _{j}^{\ast }e^{i\theta _{j}}+A_{2}(\sigma
_{j}+\sigma _{j}^{\ast }) ,  \notag \\
\frac{d\sigma _{zj}}{d\tau } &=&2\rho \left[ \left( A_{1}^{\ast }e^{-i\theta
_{j}}+A_{2}\right) \sigma _{j}+\sigma _{j}^{\ast }\left( A_{1}e^{i\theta
_{j}}+A_{2}\right) \right] -\Gamma \left( \sigma _{zj}-1\right) , \\
\frac{d\sigma _{j}}{d\tau } &=&i\left( \Delta _{20}+\frac{p_{j}}{2}\right)
\sigma _{j}-\rho \sigma _{zj}\left( A_{1}e^{i\theta _{j}}+A_{2}\right)
-\Gamma \sigma _{j} ,  \notag \\
\frac{dA_{1}}{d\tau } &=&i\Delta _{21}A_{1}+\frac{1}{N}\sum_{j=1}^{N}\sigma
_{j}e^{-i\theta _{j}}-\kappa A_1 ,  \notag
\end{eqnarray}%
where the atomic polarization and population damping $\Gamma $ as
well as the probe field damping $\kappa$ have been
phenomenologically added. The strong pump field $A_2$ is treated
as a real undepleted field and removed from Eq.(\ref{eq2}) as an
external condition \cite{Boni}. Some dimensionless variables or
transformations have been introduced in above derivations, i.e.,
$p_{j}= \frac{\left\langle \widehat{p}_{j}\right\rangle }{\hbar
k\rho }$, $\theta _{j}=2k\left\langle \widehat{z}_{j}\right\rangle
$, $A_{i}=\frac{ \left\langle \widehat{a}_{i}\right\rangle
}{\sqrt{N\rho }}e^{i\omega _{2}t}$ ($i=1,2$), $\sigma
_{j}=\left\langle \widehat{\sigma }_{j}^{-}e^{ik_{2}
\widehat{z}_{j}}\right\rangle e^{i\omega _{2}t}$ , $\sigma
_{zj}=-2\left\langle \widehat{\sigma }_{zj}\right\rangle $,
$\Delta _{21}=(\omega _{2}-\omega _{1})/(\omega _{r}\rho )$,
$\Delta _{20}=(\omega _{2}-\omega _{0})/(\omega _{r}\rho )$,
$\nu=\nu_z/(\omega_r \rho)$ and $\tau =\omega _{r}\rho t$ under
the assumption $k_{1}\thickapprox k_{2}=k$ and $ g_{1}\thickapprox
g_{2}=g$, for simplicity. The scaling parameters $\omega _{r}$ and
$\rho $ used above are two-photon recoil frequency $\omega
_{r}=2\hbar k^{2}/m$ and the dimensionless CARL parameter $\rho
=\left( \frac{g\sqrt{N}}{\omega _{r}}\right) ^{2/3}$,
respectively. It should be noted that Eq.(\ref{eq2}) are identical
to those derived in the Ref.\cite{Boni}\cite{Perrin}\cite{Kruse}
except for one single term $-\nu ^{2}\theta _{j}$ for the harmonic
trapping force, and all the simulations throughout this paper will
be based on these equations.

In order to analyze the unique discrete gain profile in Fig.2, we
focus on the long-time evolution of $A_{1}$ when the probe damping
is weak so as to be enclosed in a high-finesse ring cavity
\cite{11}. The gain quantity depicted in Fig.2 is defined by
$G\left( \Delta _{21},\tau \right) =\frac{\left\vert A_{1}\left(
\tau \right) \right\vert ^{2}-\left\vert A_{1}\left( 0\right)
\right\vert ^{2}}{\left\vert A_{1}\left( 0\right) \right\vert
^{2}}$ with $ A_{1}\left( 0\right) $ being the initial value of
the weak probe field. By formally integrating Eq.(\ref{eq2}), the
probe field can be expressed as
\begin{equation}
A_{1}\left( \tau \right) =A_{1}(0)e^{(i\Delta _{21}-\kappa )\tau
}+e^{(i\Delta _{21}-\kappa )\tau }\int_{0}^{\tau }\left(
\frac{1}{N} \sum_{j}\sigma _{j}\left( \tau ^{\prime }\right)
e^{-i\theta _{j}(\tau ^{\prime })}\right) e^{(\kappa -i\Delta
_{21})\tau ^{\prime }}d\tau ^{\prime }\text{,}  \label{eq3}
\end{equation}%
and the probe gain
\begin{equation}
G(\kappa ,\Delta _{21},\tau )=e^{-2\kappa \tau }\left[ \left\vert
\frac{ \widetilde{C}(\kappa -i\Delta _{21},\tau
)}{A_{1}(0)}\right\vert ^{2}+2Re\left( \frac{\widetilde{C}(\kappa
-i\Delta _{21},\tau )}{A_{1}(0)} \right) \right] +\left(
e^{-2\kappa \tau }-1\right) , \label{Gain}
\end{equation}%
where $\widetilde{C}(\kappa -i\Delta _{21},\tau )$ is defined by
\begin{equation}
\widetilde{C}(\kappa -i\Delta _{21},\tau )=\int_{0}^{\tau }C(\tau ^{\prime
})e^{(\kappa -i\Delta _{21})\tau ^{\prime }}d\tau ^{\prime }  \label{Laplace}
\end{equation}%
with $C(\tau )=\frac{1}{N}\sum_{j}\sigma _{j}\left( \tau \right)
e^{-i\theta _{j}(\tau )}$ being defined as a collective atomic
coherence parameter by Ref.\cite {Perrin}. For an enough long time
of evolution, provided $\tau \gg 2\pi /\nu $ with $\nu \geq
\frac{2\rho |A_{1}A_{2}|}{\Delta _{20}}$ \cite{22}, all the atoms
in the trap will approximately perform a harmonic oscillation with
different amplitudes $\theta _{j0}(\tau )$ and phases $\phi
_{j0}(\tau ) $, described by $\theta _{j}(\tau )\thickapprox
\theta _{j0}(\tau )\cos \left[ \nu \tau +\phi _{j0}(\tau )\right]
$. Then $e^{-i\theta _{j}(\tau
^{\prime })}$ in Eq.(\ref{Laplace}) will be expanded as 
\begin{equation}
\widetilde{C}(\kappa -i\Delta _{21},\tau )\approx \sum_{n=-\infty
}^{\infty } \left[ \frac{1}{N}\sum_{j=1}^{N}J_{n}\left[ \theta
_{j0}(\tau )\right] e^{in[\phi _{j0}(\tau )-\frac{\pi
}{2}]}\int_{0}^{\tau }d\tau ^{\prime }\sigma _{j}\left( \tau
^{\prime }\right) e^{\left[ \kappa -i\left( \Delta _{21}-n\cdot
\nu \right) \right] \tau ^{\prime }}\right] \text{,} \label{eq4}
\end{equation}%
where the function $J_{n}\left[ \theta _{j0}(\tau )\right] $ is
the $n$-th order Bessel function of the first kind. In above
derivations, we have assumed that the amplitude $\theta _{j0}(\tau
^{\prime })$ and phase $\phi _{j0}(\tau ^{\prime })$ are slowly
varying quantities after a long time evolution and can be moved
out from the integral of Eq.(\ref{eq4}). If we extend the upper
integral limit of Eq.(\ref{eq4}) to infinity ($\tau \rightarrow
\infty $), the integral defines a Laplace transformation of
$\sigma _{j}\left( \tau ^{\prime }\right) $, i.e.,
$\mathcal{L}\left[ \sigma _{j}\left( \tau ^{\prime }\right)
\right] =\int_{o}^{\infty }d\tau ^{\prime }\sigma _{j}\left( \tau
^{\prime }\right) \cdot $ $e^{\left[ \kappa -i\left( \Delta
_{21}-n\cdot \nu \right) \right] \tau ^{\prime }}$, on condition
that $|\sigma _{j}\left( \tau ^{\prime }\right) |$ doesn't
increase much more than $e^{\Gamma \tau }$. Generally $\sigma
_{j}\left( \tau \right) $ is limited by Bloch sphere of two-level
atom and can be approximately expanded by a Fourier series with
$\nu $ as a fundamental frequency
\begin{equation}
\sigma _{j}\left( \tau \right) \approx \langle \sigma _{j}(\tau )\rangle
+\sum_{n=1}^{\infty }c_{n}e^{in\nu t},  \label{Fourier}
\end{equation}%
where $c_{n}$ is the Fourier amplitude of the $n$-th order harmonic
components. Furthermore, if the pump-atom detuning satisfies $\Delta
_{20}\gg \Gamma $, the long-time average polarization $\langle \sigma
_{j}(\tau )\rangle $ will be
\begin{equation}
\langle \sigma _{j}\left( \tau \right) \rangle \approx
S_{0}=-\frac{\Omega \Gamma }{2\left( \Omega ^{2}+\Gamma
^{2}+\Delta _{20}^{2}\right) }-i\frac{ \Omega \Delta
_{20}}{2\left( \Omega ^{2}+\Gamma ^{2}+\Delta _{20}^{2}\right)
}\text{,}  \label{eq5}
\end{equation}%
where $\Omega =2\rho A_{2}$ is the Rabi frequency of the pump
field and $ S_{0}$ can be determined by taking an adiabatic
approximation of atomic internal freedom as done in
\cite{adiabatic}. If we only take the dc term of $\sigma
_{j}\left( \tau \right) $ in Eq.(\ref{Fourier}) for a qualitative
spectrum analysis, Eq.(\ref{eq4}) reads 
\begin{eqnarray}
\widetilde{C}(\kappa -i\Delta _{21},\tau ) &\approx
&S_{0}\sum_{n=-\infty }^{\infty }\left[
\frac{1}{N}\sum_{j=1}^{N}J_{n}\left[ \theta _{j0}(\tau ) \right]
e^{in[\phi _{j0}(\tau )-\frac{\pi }{2}]}\int_{0}^{\tau }d\tau
^{\prime }e^{\left[ \kappa -i\left( \Delta _{21}-n\cdot \nu
\right) \right]
\tau ^{\prime }}\right]   \notag \\
&=&S_{0}\sum_{n=-\infty }^{\infty
}\frac{1}{N}\sum_{j=1}^{N}J_{n}\left[ \theta _{j0}(\tau )\right]
e^{in[\phi _{j0}(\tau )-\frac{\pi }{2}]}\frac{ e^{\kappa \tau
-i\left( \Delta _{21}-n\cdot \nu \right) \tau }-1}{\kappa -i\left(
\Delta _{21}-n\nu \right) }.  \label{eq6}
\end{eqnarray}%
As the probe-field damping rate $\kappa >0$, we have $\lim_{\Delta
_{21}-n\nu \longrightarrow 0}\frac{e^{\kappa \tau }e^{-i\left(
\Delta _{21}-n\cdot \nu \right) \tau }-1}{\kappa -i\left( \Delta
_{21}-n\nu \right) }=\frac{e^{\kappa \tau }-1}{\kappa }$, then the
gain under resonant condition is
\begin{equation}
 G(\kappa ,\tau )=\left( \frac{1-e^{-\kappa \tau
}}{\kappa }\right) ^{2}\left\vert \frac{C_{0}(\tau
)}{A_{1}(0)}\right\vert ^{2}+2e^{-\kappa \tau }\frac{1-e^{-\kappa
\tau }}{\kappa }Re\left( \frac{C_{0}(\tau )}{ A_{1}(0)}\right)
+\left( e^{-2\kappa \tau }-1\right) ,  \label{Gainlimit}
\end{equation}%
where $C_{0}(\tau )=\frac{1}{N}\sum_{j=1}^{N}S_{0}e^{i\theta
_{j0}(\tau )\cos \left[ \phi _{j0}(\tau )\right] }$ is the
asymptotic coherence parameter which is determined by the secular
average atomic polarization $S_{0}$ and the instantaneous atomic
distribution at $\tau$. The coefficient $(1-e^{-\kappa
\tau})/\kappa$ of Eq.(\ref{Gainlimit}) shows here that when the
damping rate $\kappa $ is small enough and the evolution time
$\tau \gg 1/\kappa $, the intensity gain of the probe field will
be extremely high in the neighborhood of $\Delta _{21}=n\nu $, and
thus results in a uniform discrete gain structure as shown in
Fig.2 simulated directly on Eq.(\ref{eq2}). Although the Raman
gain condition will not be changed if the higher harmonic
components of Eq.(\ref{Fourier}) is included, except for a more
complicated sum formula for Eq.(\ref{eq6}), the height of the gain
peak for a specific $n$ in Fig.2 can't be interpreted by above
analytical method. As shown by Fig.2, for example, the light
amplification mostly for positive multiple of the trap frequency
and approximately up to $n=10$ is still a problem. The further
simulations only indicate that the light amplification for
positive detuning is a feature owing to the red-detuned of the
pump field from the atom frequency. However, this regular gain
spectrum still can be developed as an efficient detection method
to analyze the micro-trap potential formed by a local interference
field.

\section{Self-organization in phase space}

Under the Raman resonant condition $\Delta _{21}=n\nu $, another
ingredient that enhances the maximum gain of probe field is the
synchronization of the atomic motion in the harmonic trap, which
is indicated by the term of $C_{0}(\tau )/A_{1}(0)$ in
Eq.(\ref{Gainlimit}) and can be demonstrated through a bunching of
the atomic trajectories in the phase space. To some extent, the
present problem described by Eq.(\ref{eq2}) under the adiabatic
approximation is similar to the famous Kuramoto model of ensembles
of coupled oscillators \cite{Kuramoto} , in which the complex
order parameter $R(\tau )e^{i\Phi (\tau )}=\frac{1}{N}
\sum_{j=1}^{N}e^{-i\theta _{j}(\tau )}$ plays an equivalent role
as that of the bunching parameter $b$ \cite{Boni} in the
traditional CARL system \cite {Java}. For the complex order
parameter, $R(\tau )$ measures the phase coherence of all the
oscillators and $\Phi (\tau )$ describes their average phase. A
main result of Kuramoto model is: When the coupling intensity (in
present model the coupling intensity is related to the Rabi
frequency, $\rho A_{2}$, of the pump field) exceeds a certain
threshold, the incoherent oscillators are mutually synchronized
and $R(\tau )$ grows exponentially. In present case, when the
resonant conditions $\Delta _{21}=n\cdot \nu $ and $\Delta
_{20}\gg \Gamma $ are both satisfied, the long-time output gain of
$A_{1}$
reduces to%
\begin{equation}
G(\kappa ,\tau )\approx \left( \frac{1-e^{-\kappa \tau }}{\kappa
}\right) ^{2}\left\vert \frac{S_{0}}{A_{1}(0)}\right\vert
^{2}R_{0}^{2}(\tau )+2e^{-\kappa \tau }\frac{1-e^{-\kappa \tau
}}{\kappa }Re\left( \frac{ S_{0}e^{i\Phi _{0}(\tau
)}}{A_{1}(0)}\right) R_{0}(\tau )+\left( e^{-2\kappa \tau
}-1\right) ,  \label{gain0}
\end{equation}
where $R_{0}(\tau )e^{i\Phi _{0}(\tau
)}=\frac{1}{N}\sum_{j=1}^{N}e^{i\theta _{j0}(\tau )\cos \left[
\phi _{j0}(\tau )\right] }$ is the secular complex order parameter
determined only by the atomic distribution in the phase space.
Eq.(\ref{gain0}) indicates that the field gain is explicitly
dependent on the increase of collective atomic coherence, $R(\tau
)$.
\begin{figure}[htp]%
\includegraphics[height=6cm,width=8cm]{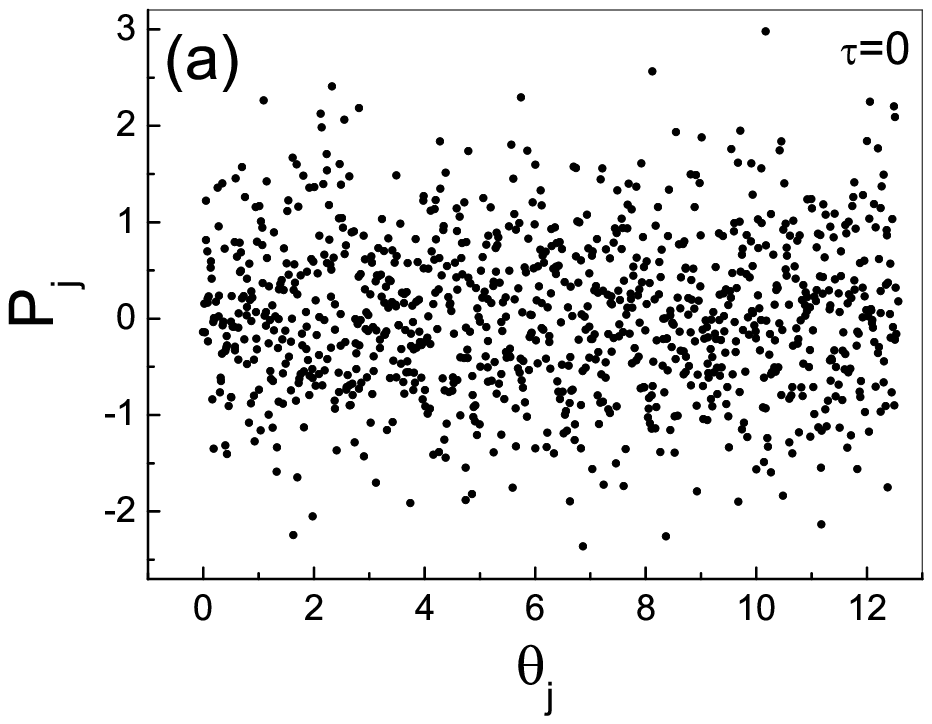}
\includegraphics[height=6cm,width=8cm]{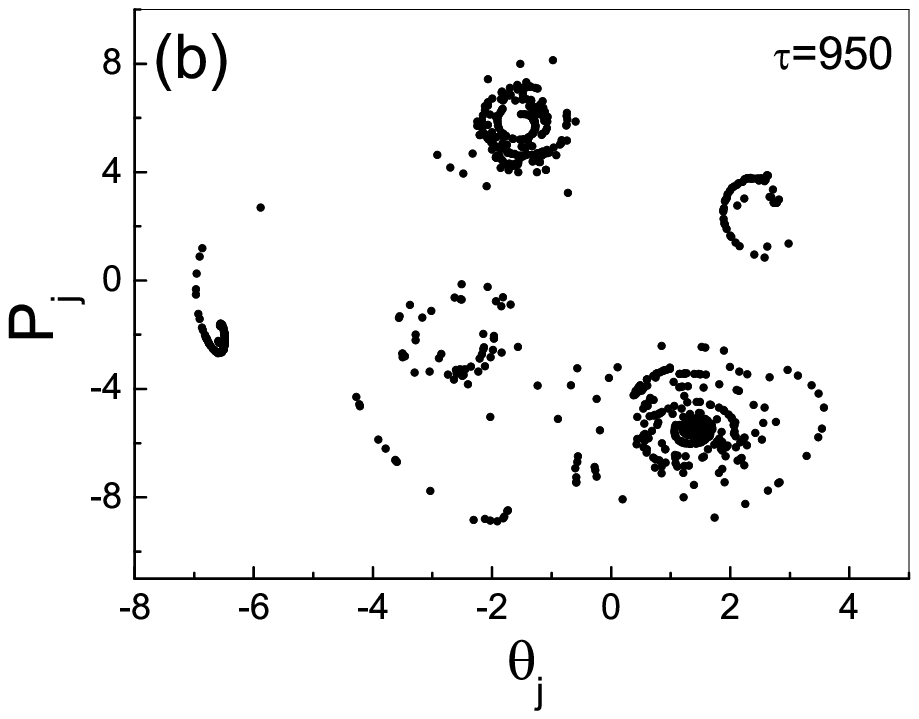}
\includegraphics[height=6cm,width=8cm]{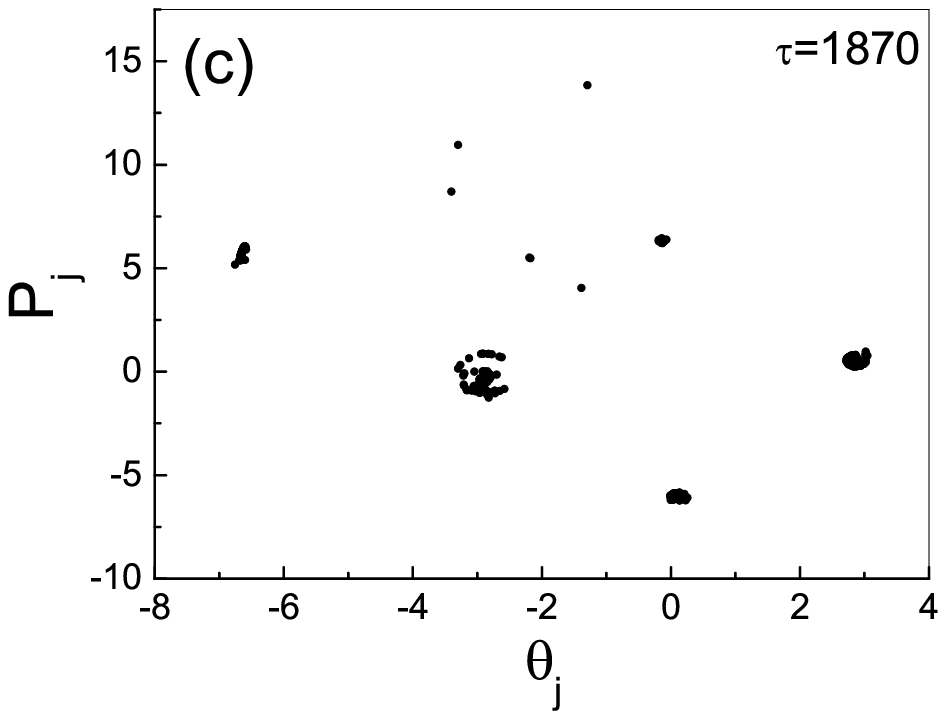}
\includegraphics[height=6cm,width=8cm]{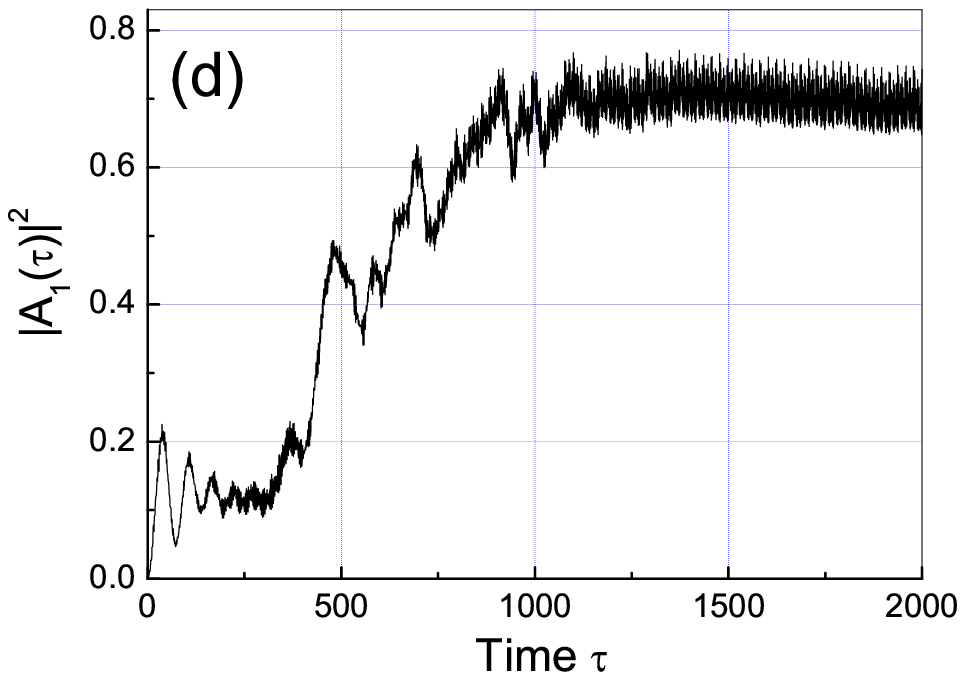}
\caption{Sample snapshot distributions of $1000$ atoms in phase
space (a) at $\tau=0$ (Initial distribution), (b) at $\tau=950$,
(c) at $\tau=1800$ and (d) the corresponding time evolution of the
probe intensity $|A_1(\protect\tau)|^2$ under the resonant
condition with parameters $\protect\Delta_{21}=2\nu$, $\protect\nu
=2$, $\Gamma =1, \protect\kappa=0.01$, $\protect\rho =3$, $A_2=2$,
and $\Delta _{20}=-15$.}
\end{figure}%
At an early stage of evolution, as the atoms incoherently scatter
the pumping photon into probe mode and their spatial distribution
in the trap almost keeps uniform with $R(\tau )\sim 0$ (Fig.3(a)),
the probe field fluctuates around a small value. After a period of
lethargy, a self-organization structure is established through the
recoil motion modified by the trap and the atoms in phase space
concentrate into several clusters (Fig.3(b)), resulting in an
exponential growth of the probe field. When this process is
saturated (Fig.3(c)) by a counteraction of the spontaneous damping
processes \cite{Robb1}, the coherent parameter reaches its maximum
value, or extremely $ R(\tau )\rightarrow 1$ (all the atoms moving
synchronously as one big atom), and a stable output of $A_{1}$
will be set up. Fig.3(d), which is directly simulated from
Eq.(\ref{eq2}), displays a typical amplification process of the
probe intensity under this resonant condition and, finally,
establishes a saturated output with an efficiency nearly to $18\%$
(Here the efficiency is defined by the intensity rate between
pumping field and the output probe field). According to a further
Fourier analysis, the fast oscillation observed in Fig.3(d) at
saturation is induced by the harmonic motion of the atoms in the
trap.
\begin{figure}[htp]
\includegraphics[scale=0.75]{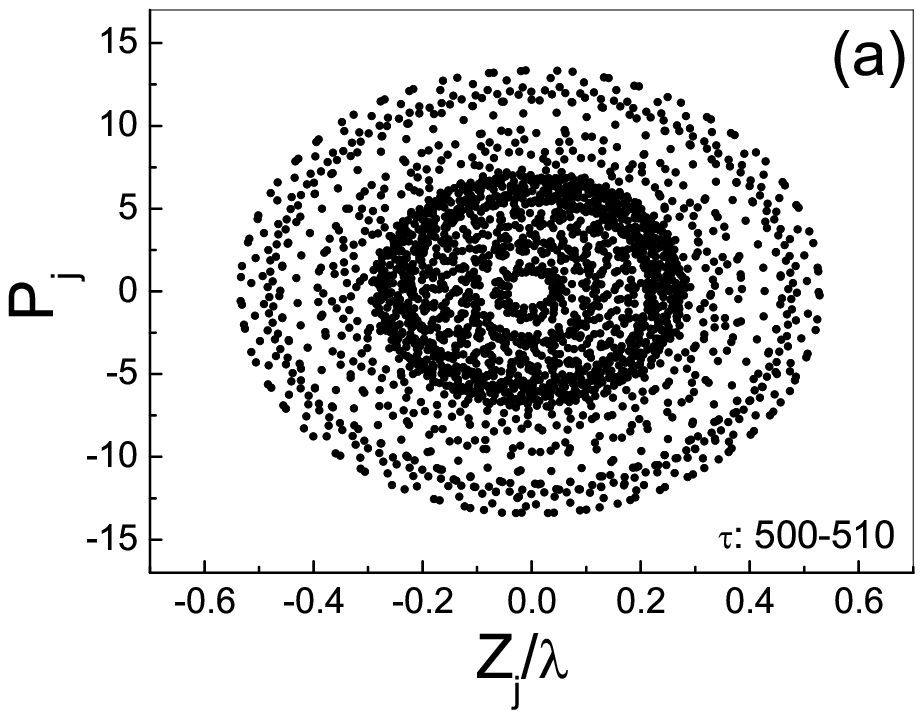}
\includegraphics[scale=0.75]{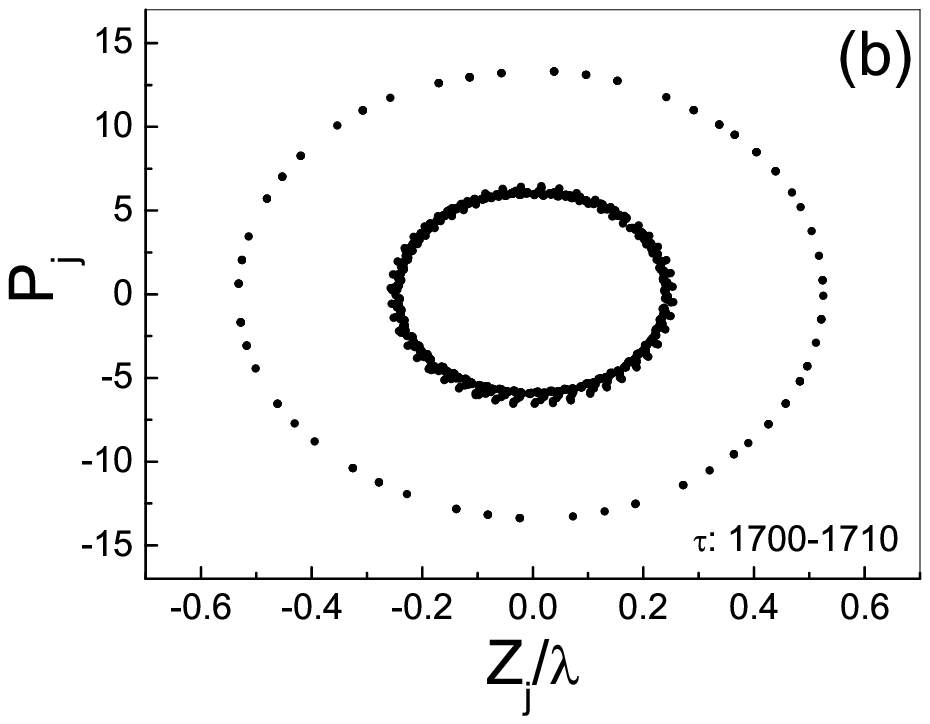}
\caption{ The atomic trajectories in phase space during (a)
$500<\tau<510$ and (b) $1700<\tau<1710$ with parameters
$\protect\nu =2$, $\Gamma =1, \protect\kappa=0.01$, $\protect\rho
=3$, $A_2=2$, $\Delta _{20}=-15$ and $\Delta_{21}=4$.}
\end{figure}

During the lasing process, the self-organization structure related
to the light amplification corresponds to a spontaneous bunching
of atomic trajectories in the phase space (Fig.4), which,
actually, is similar to the atomic density grating in the free
space CARL system when $b$ is large. In order to demonstrate the
spatial features of this atomic self-organization behavior, the
initial spatial distribution of the atoms are set uniform in a
region of $[0,\lambda ]$ and the initial atomic momentum follows a
Gaussian distribution (showed by Fig.3(a)) with a thermal spread
of width above recoil temperature (about $0.8\rho\hbar k $)
\cite{semi}. Before the increment of $A_1$, the atomic
trajectories are almost ubiquitous in the phase space and, after a
stable output of $A_{1}$ is established, most atomic trajectories
converge into the antinode area (around $0.25\lambda $) and the
other ones into the node area (around $0.5\lambda $) of a standing
wave induced by an interference between the amplified probe field
and the strong pump field, with its wavelength being defined by
$\lambda =4\pi /(k_{1}+k_{2})$. The bunching positions of the most
atomic trajectories into the antinode area are a clear result due
to the red-detuned of the pump field $\Delta _{20}=-15 <0$ in
Fig.4.

According to more simulations, we find that the amplification and
the pattern formation of the atoms in phase space under the Raman
resonant condition are robust to other parametric values provided
that the gas medium is operated on a far-off red-detuned region
and the trap is strong enough \cite{22}. Following above
conditions the change of collective coupling constant $\rho$ in
the resonant gain region will not qualitatively change the
self-organization effect and the long time high CW gain output of
this model.

\section{ Discussions}

The experimental realization of above CW CARL system can be
carried out by loading the laser-cooled atoms into an overall
magnetic trap or into a far-off detuned optical lattice.
Thereafter, the trap is displaced along its axis for about one
light wavelength $\lambda$ to trigger the atomic motion, and then
the atoms are pumped by a strong red-detuned field. A gain of a
seed probe field in the backward direction under the resonant
conditions $\Delta _{21}=0,\pm \nu ,\pm 2\nu ,\pm 3\nu \cdots $
will be detected then. This process will be enhanced further by a
collective behavior of the atoms due to the synchronization of
their recoil motion under Raman resonant condition. The strong
pump field feeds the energy to the system (excite the trapping
oscillating mode by atomic recoil motion) and the dipole force,
generated by the interference of the probe and pump light, bunches
the atoms into several groups to adjust them synchronically
extracting kinetic energy from the trap mode into the probe field
\cite{Domokos}. This scheme will also present a clue to introduce
a new configuration of free electron lasers (FEL) where the high
speed electrons are injected into a (or an array of) strong
quantum trap (e.g. quantum dots). By stimulating the electrons'
motion through electromagnetic field on the electrode with
appropriate frequency, the high gain field with certain frequency
will be generated when the electrons oscillate collectively in (or
through) these quantum traps.

In conclusion, we have investigated the light amplification in a
CARL with the atomic center-of-mass motion influenced by an
external harmonic potential. Under appropriate parametric
condition that the pump field is red-detuned from the atomic
transition frequency, our results show that CARL confined by a
robust trap can give rise to a CW laser with a higher lasing gain
than that of the CARL in the free space. The optical gain profile
as a function of the pump-probe detuning displays a discrete
structure and another kind of atomic self-organization related to
the light gain is demonstrated in the phase space. This work
suggests a new way to promote a CW CARL device with a higher gain
and, perhaps, analogically implies a new configuration of FEL.

\section*{Acknowledgement}

We would like to thank Dr. O.Zobay for helpful discussions. The
work is supported by the Science Foundation of the Education
Bureau of Shaanxi province under grant No.05JK140 and by the key
project No.ZK2406 of Baoji University of Arts \& Science.


\begin{thebibliography}{99}
\bibitem{22} The condition $\nu \geq \frac{2\rho |A_1 A_2|}{\Delta_{20}}
$ means that the trap potential $\nu$ is greater than the dipole
potential $ \frac{2\rho |A_1 A_2|}{\Delta_{20}}$ induced by the
intereference of probe field and the pump field.
\bibitem{Robb1} The heating processes due
to friction and diffusion effects are not considered in this
model. The problem on this topic can see \cite{Robb}.
\bibitem{semi} This condition is for the
validity of the semiclassical approximation of this model.
\end{thebibliography}
\end{document}